\pdfoutput=1
\documentclass[aps,pre,twocolumn,floatfix,reprint]{revtex4}

\usepackage{color}
\usepackage{graphicx}
\usepackage{amssymb,amsfonts,amsmath}
\usepackage{hyperref}
   
\begin{document}

\title{Manifestations of Dynamical Facilitation in Glassy Materials}

\author{Yael S. Elmatad} 
\thanks{These authors contributed equally to this work.}
\affiliation{Center for Soft Matter Research, Department of Physics, New York University, New York NY, USA}
\author{Aaron S. Keys$^*$} 
\email[Corresponding author.  E-mail: ]{askeys@umich.edu}
\affiliation{Department of Chemistry, University of California, Berkeley CA, 94720}
\affiliation{Lawrence Berkeley National Laboratory, Berkeley CA, 94720}

\date{\today}

\begin{abstract}

By characterizing the dynamics of idealized lattice models with a tunable kinetic constraint, we explore the different ways in which dynamical facilitation manifests itself within the local dynamics of  glassy materials.  Dynamical facilitation is characterized both by a mobility transfer function, the propensity for highly-mobile regions to arise near regions that were previously mobile, and by a facilitation volume, the effect of an initial dynamical event on subsequent dynamics within a region surrounding it.  Sustained bursts of dynamical activity -- avalanches -- are shown to occur in kinetically constrained models, but, contrary to recent claims, we find that the decreasing spatiotemporal extent of avalanches with increased supercooling previously observed in granular experiments does not imply diminishing facilitation.  Viewed within the context of existing simulation and experimental evidence, our findings show that dynamical facilitation plays a significant role in the dynamics of systems investigated over the range of state points accessible to molecular simulations and granular experiments.

\end{abstract}

\maketitle

\section{Introduction}

When glassy materials, such as supercooled liquids, dense colloidal suspensions or driven granular materials, are cooled or compressed towards the glass or jamming transitions, the motions of their constituent particles become increasingly correlated in space and time~\cite{ediger2000spatially, weeks2000three, keys2007measurement}.  This phenomenon, known as dynamical heterogeneity (DH), is a universal property of glassy materials~\cite{glotzer2000spatially, kegel2000direct, dauchot2005dynamical} and is thought to have a direct connection with the anomalous transport properties of these systems near the glass and jamming transitions~\cite{adam1965temperature, xia2001microscopic, garrahan2002geometrical, garrahan2003coarse, berthier2011theoretical, debenedetti2001supercooled, liu1998jamming, biroli2007jamming}.  

Despite its ubiquity, the microscopic mechanism of DH remains uncertain.  Recent insights into this problem have demonstrated that DH can be decomposed into smaller dynamical sub-units.  This was first discovered by Glotzer and co-workers~\cite{gebremichael2004particle, vogel2004particle}, who showed that strings of highly-mobile particles~\cite{donati1998stringlike} are made up of shorter micro-strings whose character does not change with temperature, although the strings themselves grow with supercooling.  More recently, Candelier, Dauchot, Biroli and co-workers showed that, for both granular materials~\cite{candelier2009building, candelier2010dynamical} and molecular simulations ~\cite{candelier2010spatiotemporal}, clusters of particles undergoing nearly-simulataneous cage escapes coalesce into larger mobile clusters.  A recent comprehensive simulation study showed that DH builds up from localized excitation dynamics involving the collective displacements of only a handful of neighboring particles spanning a few molecular diameters~\cite{keys2011excitations}.  All of these observations imply the existence of a degree of spatiotemporal correlations between dynamical subunits, where dynamical events facilitate subsequent dynamics nearby in space, giving rise to large-scale DH over time~\cite{garrahan2002geometrical, garrahan2003coarse}.  The extent to which such dynamical facilitation (DF) plays a role in the relaxation mechanism of glassy materials is disputed, even for systems for which the local dynamics can be resolved directly.  This lack of agreement stems from uncertainty regarding how DF manifests itself on the microscopic level, and as a result, several different methods have been employed for measuring DF, leading to contrasting interpretations.  In particular, several studies~\cite{vogel2004spatially, bergroth2005examination, keys2011excitations} report that DF is present at all supercooled state points, whereas others argue that DF must be augmented by another structural relaxation mechanism~\cite{candelier2010spatiotemporal} or is only relevant over a narrow range of state points~\cite{candelier2010dynamical}.

Here, we apply several dynamical characterization schemes, originally proposed in the context of measuring DF in molecular simulations and granular experiments, to kinetically constrained lattice models~\cite{ritort2003glassy}, for which DF is the primary relaxation mechanism by construction.  These quantities are also measured for hybrid models that range between a hard dynamical constraint (pure facilitation) and a non-interacting lattice gas, which allows us to study the effect of delocalized soft rearrangements in violation of DF of the type proposed by Ref.~\cite{candelier2010spatiotemporal}.  To allow for comparison with molecular simulations and granular materials, we formulate a displacement field for kinetically constrained models that is analogous to coarse-grained particle displacements within a small region of space for particulate systems~\cite{keys2011excitations}.  We show that mobility transfer correlations, based on the exchange of mobility amongst neighboring spatial regions, and facilitation volumes, based on correlations between dynamical events and subsequent dynamics within the surrounding sub-volume, accurately reflect the degree of DF within a given system.  We verify that avalanches of sustained dynamical activity arise from facilitated dynamics, as predicted by Refs.~\cite{candelier2009building, candelier2010spatiotemporal, candelier2010dynamical}, but in contrast with the interpretations of Ref~\cite{candelier2010dynamical}, we find that the tendency for avalanches to contain fewer dynamical subunits with increased supercooling does not imply diminishing facilitation.   We predict that a decreasing role of DF could be detected by mobility transfer correlations that go through a maximum, or facilitation volumes that exhibit little growth with increased supercooling, although these behaviors have not been observed in either simulations or experiments.

\section {Models and Simulations}

\begin{figure*}[ht!]
\centerline{\includegraphics[width=0.85\textwidth]{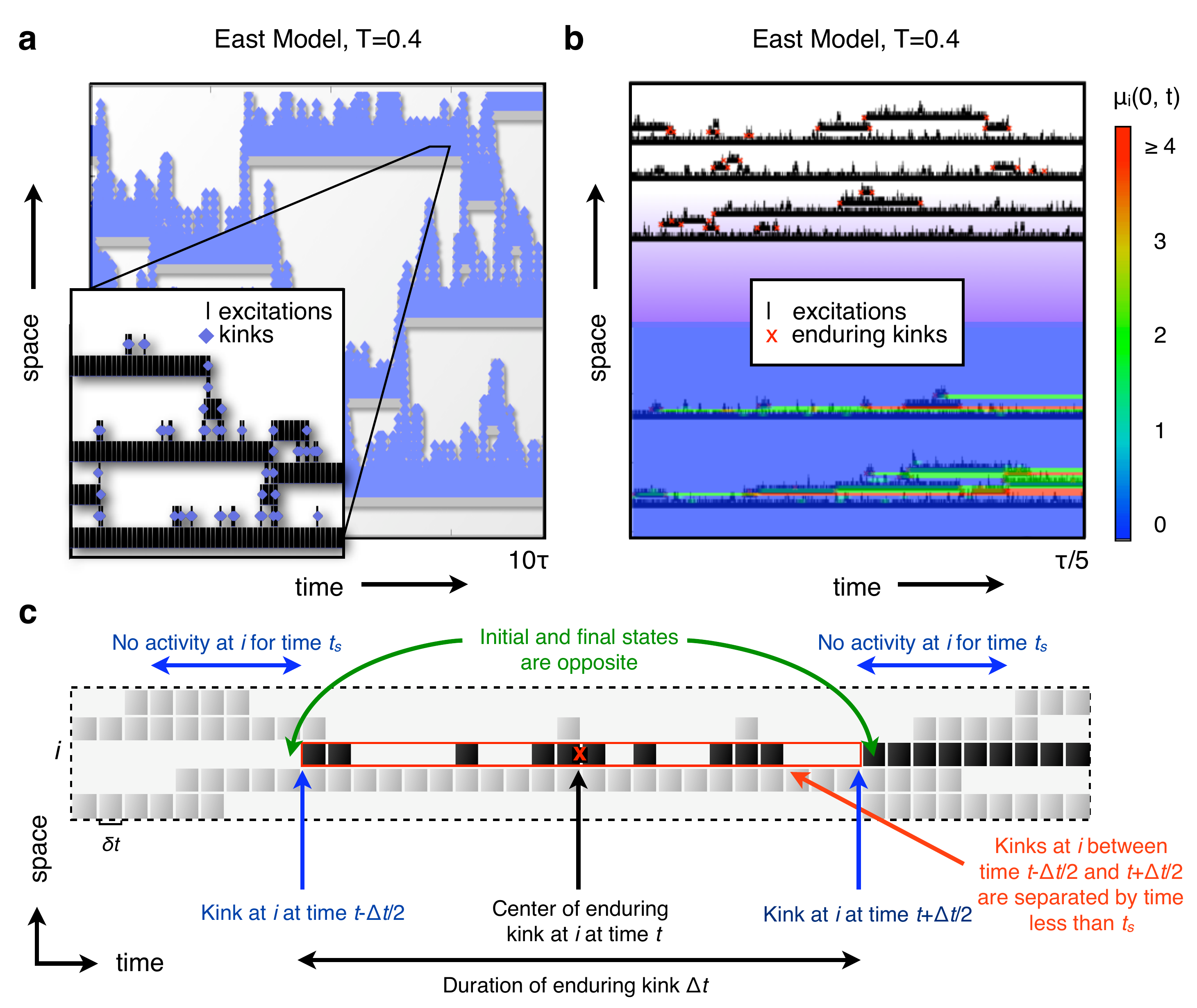}}
\caption{\label{fig:fig1} {Dynamics of kinetically constrained models.  (a) The panel depicts excitations and kinks for a long trajectory in the east model.  While the excitations connect throughout space and time, the kinks are temporally separated.  The inset illustrates that this feature becomes apparent on shorter time scales. (b)  The panel depicts a short trajectory for the east model.  Enduring kinks (long-lived changes in the arrangement of excitations) are depicted by red x's.  At the bottom of the panel, lattice sites are colored according to $\mu_i(0, t)$, the number of enduring kinks at site $i$ up to the time $t$.   (c) The schematic depicts a small region of space-time, containing cells that span an elementary time step $\delta t$, with dark shaded cells containing excitations, $n_j(t)=1$, and white cells containing no excitations, $n_j(t) = 0$.  Black cells highlight excitations at a lattice site $i$, which exhibits an enduring kink at time $t$, denoted by a red x.  Enduring kinks are defined by the binary indicator function $h_i(t)$, which itself is defined in terms of three binary operators.  These operators, $\mathcal{S}_i(t,\Delta t)$, $\mathcal{T}(t,\Delta t)$, and $\mathcal{O}(t,\Delta t)$, are described pictorially by arrows and text in the panel, and color-coded blue, red, and green, respectively.  Corresponding mathematical definitions are provided in the text.}}
\end{figure*}

DF presumes that the glassy material contains localized soft-spots, or excitations, that allow for local structural rearrangement.  These rearrangements facilitate the birth and death of excitations nearby in space, thereby facilitating nearby motion at a later time.  This physical picture is encoded into a class of kinetically constrained models (KCMs) that represent excitations as spins in a non-interacting lattice gas, where lattice sites change state only in the presence of neighboring excitations.  As a result of this dynamical constraint, these simple models exhibit the complex structural relaxation behavior of glassy materials~\cite{garrahan2003coarse, elmatad2009corresponding, xu2009equivalence}, although their equilibrium thermodynamic behavior is trivial.

We consider one-dimensional ($d$=1) systems with $N=1024$ lattice sites occupying one of two states $n_{i} = \{0, 1\}$, where 1 (0) represents an excited (unexcited) state.  The system Hamiltonian is given by $\mathcal{H} = \sum_i n_i$, and thus excitations are present with equilibrium concentration $c = \langle n_{i} \rangle = (1+e^{\beta})^{-1}$.  The inverse temperature is given by $\beta = 1/k_{\mathrm{B}}T$ with Boltzmann's constant $k_{\mathrm{B}}$ taken as unity.  We consider systems with a directional dynamical constraint defined by the 1$d$ east model~\cite{eastmodel, ritort2003glassy}, where sites with $n_{i} = 1$ can facilitate the adjacent site $n_{i+1}$.  Structural relaxation in the east model follows from hierarchical dynamics of the type proposed by Ref.~\cite{palmer1984models} and reported in simulations of atomistic supercooled liquids~\cite{keys2011excitations}.  Below the glassy dynamics onset temperature, $T_\mathrm{o}$, this structural relaxation law is in good agreement with simulation and experiment~\cite{elmatad2009corresponding}.  We approximate $T_\mathrm{o}$ for our systems as the maximum $T$ for which the systems exhibits significant four-point correlations, $T_\mathrm{o} \approx 1$.  The structural relaxation time $\tau$ is defined as the mean time required to for relaxed regions to span the mean distance between excitations, $\ell=c^{-1/d}$.

In addition to facilitated moves, we allow for moves that violate the kinetic constraint with probability $\exp (-\beta U_\mathrm{soft})$.  The parameter $U_\mathrm{soft}$ represents the energy barrier for soft, delocalized dynamics, and is systematically varied throughout the study.  Taking into account the constraint and detailed balance, the transition rates for a given lattice site are given by $k_{0 \rightarrow 1} = \exp(-\beta)[n_{i-1}+\exp(-\beta U_\mathrm{soft})]$ and $k_{1 \rightarrow 0} = n_{i-1} + \exp (-\beta U_\mathrm{soft})$.  {The functional form for soft relaxation is chosen so as to allow for the possibility of a crossover, where at low temperatures, soft relaxation becomes more probable than facilitated dynamics, as postulated by Ref.~\cite{candelier2010dynamical}.  Other functional forms satisfying this criterion are possible, provided they exhibit a weaker temperature dependence than the super-Arrhenious relaxation law of hierarchical models.}  In addition to the directional east model, in many cases, we have compared our results with the non-directional Fredrickson-Andersen model~\cite{fredrickson1984kinetic}, and verified that they are qualitatively similar, although these results are not presented here.

\section{Dynamics of Kinetically Constrained Models}

DF is trivial to measure in KCMs because the excitations themselves are directly observable.  Such direct measurement is not currently possible in molecular simulations or related experiments, because the precursors to excitation dynamics, if they exist, are not yet known.  Instead, DF must be inferred from dynamical quantities.  Dynamics in molecular systems correspond to local particle rearrangements, which, in turn, correspond to rearrangements in the underlying positions of excitations.  Thus, by analogy, dynamics in KCMs correspond to changes in microstate, or kinks.  A kink occurs at site $i$ and time $t$ if $\kappa_i(t) = n_i(t) - n_i(t-\delta t)$ is $\pm 1$, where $\delta t$ is an elementary time step.  Although excitations connect throughout space and time, kinks become disconnected when viewed on short time scales, as illustrated by the main panel and inset of Fig.~\ref{fig:fig1}a for a long east model trajectory.   

{Molecular simulations show that particles in supercooled liquids exhibit ubiquitous high-frequency, small amplitude displacements that are often reversed, such that particles surge back and forth over relatively long periods of time before eventually sticking to new positions~\cite{keys2011excitations, chandler2010dynamics, widmer2009central}.  These surging motions are distinct from harmonic oscillations; they can be observed from time series of time-coarse-grained coordinates or inherent structures~\cite{stillinger1982hidden, heuer2008exploring}, where the instantaneous molecular configuration at each time slice is quenched to a local potential energy minimum.  Surging is also observed for the east model and other hierarchical KCMs.  In these systems, the majority of kinks are quickly reversed, giving rise to fleeting changes in the underlying configuration of excitations.  This behavior is illustrated in the inset of Fig.~\ref{fig:fig1}a.}

{Measurements of DF carried out for glass-forming liquids and granular materials are largely based upon the spatiotemporal distribution of particle displacements.  This displacement field is a more complicated quantity than the simple binary operators considered in studies of transport decoupling~\cite{jung2005dynamical, hedges2007decoupling}, where the dynamics of interest -- the presence or absence of motion -- is well described by kinks.  Deriving a field of particle displacements for KCMs requires that surging kinks be coarse-grained away, as these types of motions do not contribute to the net displacement of particles.  This coarse-graining is performed by considering a subset of ``enduring kinks'' that produce a change in the configurational state of excitations that is maintained for a significant period of time.  This time scale, denoted by a sojourn time $t_s$, is taken here as the mean time scale for dynamical exchange events, $\left< \tau_\mathrm{x} \right>$~\cite{jung2005dynamical}.  For particulate models, $\left< \tau_\mathrm{x} \right>$ is on the order of the plateau times used to characterize excitation dynamics~\cite{hedges2007decoupling, keys2011excitations}.  Dynamical events within this plateau regime are localized, involving the cooperative displacement of only a handful of neighboring particles~\cite{keys2011excitations}.  Larger particle displacements build up from smaller, more elementary events~\cite{keys2011excitations}.  Elementary excitation dynamics provide the physical basis for the enduring kinks considered here.  However, because excitation dynamics are self-similar over a range of length scales~\cite{keys2011excitations}, enduring kinks might also approximately represent similar dynamical quantities, such as micro-strings~\cite{gebremichael2004particle}, clusters of cage escapes~\cite{candelier2009building}, or rearrangements of elementary subsystems within the potential energy landscape~\cite{rehwald2010coupled}.}

{Given a time series of kinks along a trajectory, $\kappa_i(t)$, we define a set of enduring kinks occurring at site $i$ and times $t$ with durations $\Delta t$.}  An enduring kink at site $i$ and time $t$ is indicated by a function $h_i(t)$, defined below in terms of three binary operators.  These operators are described in the color-coded schematic shown in Fig. \ref{fig:fig1}c.  The first operator, $\mathcal{S}_i(t,\Delta t)$, requires that kinks on both ends of the trajectory endure longer than a sojourn time, $t_s$,
\begin{eqnarray}
\begin{split}
\mathcal{S}_i(t, \Delta t) = &  \prod_{s=\{-1, 1\}} \kappa^2_i(t + s \Delta t/2) \\
& \times \prod_{t' = \delta t}^{t_\mathrm{s}} \left[ 1 - \kappa_i^2(t + s \Delta t/2  + st') \right].
\end{split}
\label{eqn:S}
\end{eqnarray}
The first product requires that kinks occur at both ends of the enduring kink event, spanning from $t-\Delta t/2$ to $t+\Delta t/2$.  The second product ensures that no kinks occur within a time $t_s$ prior to the first kink or within a time $t_s$ after the final kink.  When these conditions are satisfied, $\mathcal{S}_i(t,\Delta t) = 1$.  Otherwise, it equals zero. 

The transient portion of the trajectory is defined such that at least one kink occurs within every sliding time window of size $t_\mathrm{s}-\delta t$ between time $t-\Delta t/2$ and $t+\Delta t/2$,
\begin{equation}
\mathcal{T}_i(t, \Delta t) = \prod_{t'=t-\Delta t /2}^{t+\Delta t/2 - t_\mathrm{s}} \left[1-\delta\left(\sum_{t''=t'+\delta t}^{t'+t_\mathrm{s}} \kappa^2_i(t'')  \right)\right].
\label{eqn:T}
\end{equation}
{Here, $\delta(\cdots)$ denotes the Kronecker delta function.}  When this criterion is satisfied, $\mathcal{T}_i(t,\Delta t) = 1$.  Otherwise, it equals zero.  The transient portion of the trajectory must contain an odd number of kinks, such that the event results in an overall change in state,
\begin{equation}
\mathcal{O}_i(t, \Delta t) = \left (\sum_{t'=t-\Delta t /2}^{t+\Delta t/2 - t_\mathrm{s}} \kappa_i(t') \right)^2.
\label{eqn:O}
\end{equation}
$\mathcal{O}_i(t,\Delta t) = 1$ when an odd number of kinks have occured between time $t-\Delta t/2$ and $t+\Delta t/2$ and zero otherwise.
The path functional for an enduring kink is then given by,
\begin{equation}
h_i(t) = \sum_{\Delta t} \mathcal{S}_i(t, \Delta t) \mathcal{T}_i(t, \Delta t) \mathcal{O}_i(t, \Delta t) .
\label{eqn:hi}
\end{equation}
The term in the summation is a product over all of the binary operators defined above.  The summation is carried out over all possible durations, $\Delta t$, where, by construction, only one value of $\Delta t$ can satisfy all of the operators simultaneously.  The function $h_i(t)$ is therefore itself a binary operator and equals unity if and only if all three conditions are satisfied.  Otherwise, $h_i(t)$ equals zero.  

The displacement at a lattice site $i$ over a time window $t'-t$ is approximated as a simple sum of enduring kinks,
\begin{equation}
\mu_i(t, t') = \sum_{t'' = t}^{t'} h_i(t'').
\end{equation}
In molecular systems, the total displacement of a given particle does not typically scale linearly with the number of discrete displacements, since it is unusual for displacements to occur in exactly the same direction.  Thus, the mobility field may be more accurately described by $\mu_i(t, t')^{p}$, where the power $p$ depends on the fractal dimensionality of particle diffusion~\cite{allegrini1999dynamic, oppelstrup2009anomalous}.  For simplicity, we assume that $p = 1$, as this does not effect the qualitative behavior of the measurements performed here.  More realistic mappings might also account for temporary displacements that arise from reversible surging.  We observe that quantities based on $\mu_i(t, t')$ exhibit artifacts for $t<t_\mathrm{s}$, due to coarse-graining on a time scale $t_\mathrm{s}$.  This does not affect the measures considered here, which involve significantly longer time scales. {We find that qualitatively similar displacement fields can be obtained from a derivation based on the probe particle picture of Ref.~\cite{jung2004excitation}.  We use enduring kinks because they are significantly cheaper computationally and seem to provide a more direct physical connection with particulate systems.}

\section{Measuring Dynamic Facilitation}

With the dynamics of KCMs defined, we study the properties of three different measures of DF, originally proposed in context of simulations of glass-forming liquids and granular materials in experiment.  The measurements are presented chronologically, and compared based on their proficiency at detecting DF, their temperature variation, and their ability to distinguish between systems with a hard dynamical constraint and those with softened dynamics. 

\subsection{Mobility Transfer Function}

\begin{figure}
\centerline{\includegraphics[width=0.85\columnwidth]{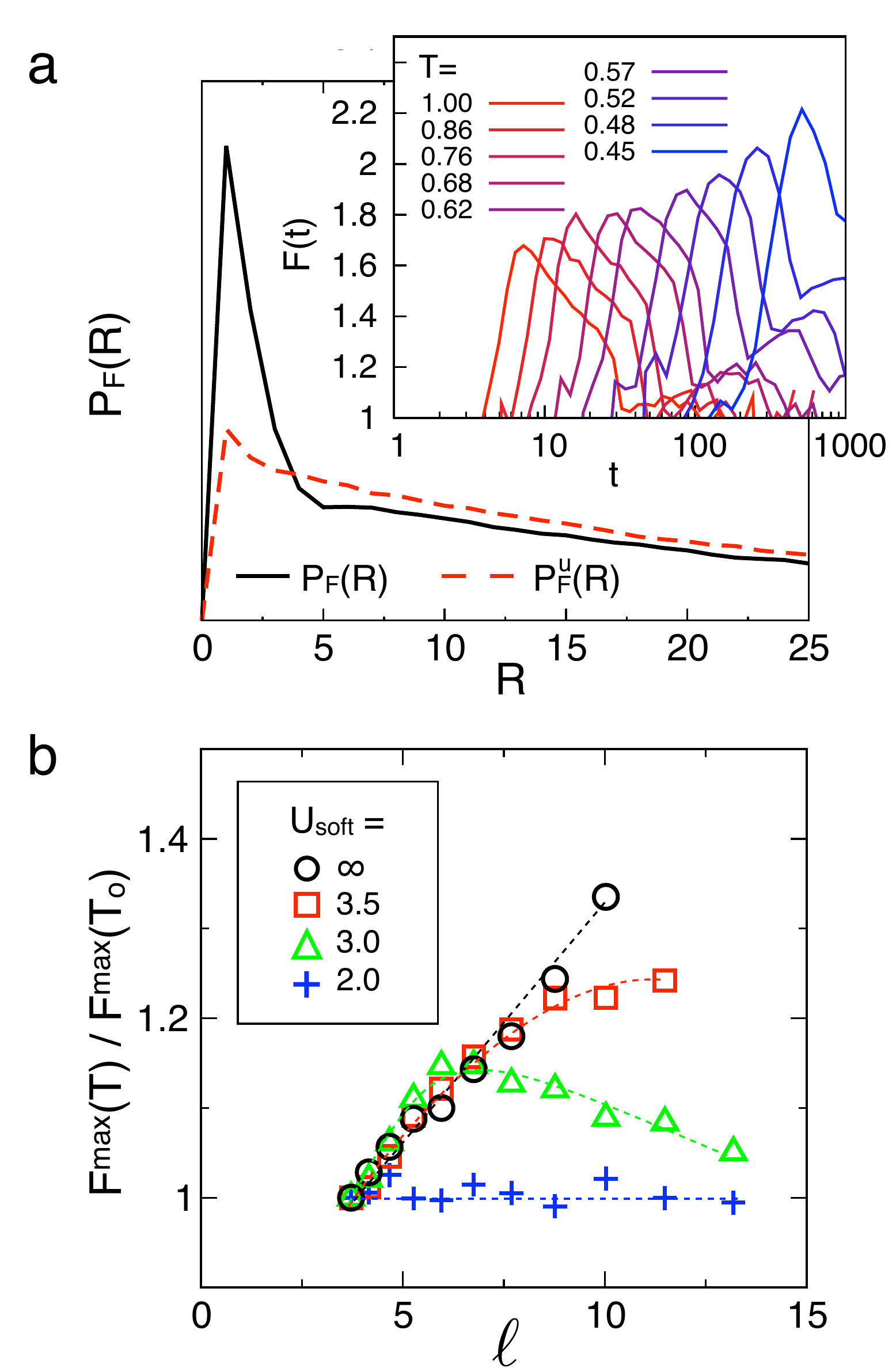}}
\caption{\label{fig:fig2}  { Mobility transfer function.  (a) The main panel shows a representative curve for $P_\mathrm{F}(R)$, the probability of observing a newly-mobile site a minimum distance $R$ away from a previously-mobile site.  {Data are shown for $T=0.45$ and $t=510$.}  The red dashed curve shows the uncorrelated result, obtained by choosing the initial ``mobile'' sites at random.  The inset shows the value of $F(t)$, the integrated difference between the curves for each $t$, which represents the degree of DF.  (b)  Comparison of the peak value of $F(t)$, $F^\mathrm{max}(T)$, as a function of temperature for different values of the softness parameter, $U_\mathrm{soft}$}}
\end{figure}

Vogel, Glotzer and co-workers have proposed a mobility transfer function based on the probability of observing new mobile particles near particles that were previously mobile.  For facilitated dynamics, this probability exceeds the uncorrelated case, and this trend becomes more apparent with supercooling, as $c$ decreases~\cite{vogel2004spatially, bergroth2005examination}.  

{Mobile lattice sites are defined in terms of the displacement field $\mu_i(t, t')$.  To ensure that no two sites exhibit exactly the same mobility, a small perturbation is introduced into the displacement field at each site,}
\begin{equation}
\mathcal{M}_i(t', t'') = \mu_i(t', t'') + \mathcal{U}(0, \epsilon).
\end{equation}
$\mathcal{U}(0, \epsilon)$ is a uniform random number on the interval $[0,  \epsilon]$, where $\epsilon$ is a small number $\ll 1$.  {Thus, in the case that several sites exhibit the same value of $\mu_i(t, t')$, the most mobile sites are chosen at random.}  We define mobile sites according to the following binary operator,
\begin{equation}
m_i(t', t'') = \Theta \left [\mathcal{M}_i(t',t'') - \mu_{\mathrm{cut}} (t',t'')\right].
\end{equation}
Here, $\Theta$, is the Heaviside step function, and $\mu_\mathrm{cut}$ is chosen to include the 5\% most mobile sites during a time window spanning from $t'$ to $t''$,
\begin{equation}
\int_{\mu'=0}^{\mu_\mathrm{cut}(t', t'')} \sum_{i=0}^N  \delta \left[\mathcal{M}_i \left(t',t'' \right) - \mu' \right] d\mu'  = \lfloor 0.95 N \rfloor.
\end{equation}
Here, $\lfloor \dots \rfloor$ denotes the floor function.  Sites with $m_i(t',t'') = 1$ comprise the highly-mobile subset.  For all other sites, $m_i(t',t'')=0$.  {The 5\% most mobile sites are chosen to correspond with molecular simulations, where cutoffs in the range 5-10\% are found to maximize the distinction between mobile and immobile particles~\cite{gebremichael2004particle, vogel2004spatially, bergroth2005examination}.  We find that all cutoffs within this range yield qualitatively similar results for the models under investigation.}

{A set of newly-mobile sites within the subsequent time window is defined by,}
\begin{equation}
w_i(t', t'') = \left[ 1-m_i(2t'-t'', t') \right] m_i(t', t''),
\end{equation}
The time interval is symmetric about $t'$.  The operator $w_i(t',t'') = 1$ for sites that are newly mobile, and zero otherwise.  {For each newly-mobile site, we measure the minimal distance $R$ to a previously-mobile site, 
\begin{equation}
 R_i = \underset{ D(i,j) }{\operatorname{arg\ min}} \left \{-\frac{1}{ D(i,j) }  m_j(t',t'') w_i(t', t'') \right \}.  
 \end{equation}
The function $D(i, j)$ is the distance between sites after accounting for periodic boundary conditions.}  The probability distribution of $R$ as a function of the time window $t = t'' - t'$, $P_\mathrm{F}(R, t)$, is compared to the uncorrelated distribution, $P_\mathrm{F}^{\mathrm{u}}(R, t)$.  For this distribution, $\mathcal{M}_i(t',t'')$ is replaced by $\mathcal{U}(0, \epsilon)$, such that the $N - \lfloor 0.95 N \rfloor$ ``mobile'' sites are chosen at random.

Both $P_\mathrm{F}(R, t)$ and $P_\mathrm{F}^{\mathrm{u}}(R, t)$ are plotted in Fig.~\ref{fig:fig2}a for a typical state point and value of $t$.  The value of $P_\mathrm{F}(R, t)$ exceeds $P_\mathrm{F}^{\mathrm{u}}(R, t)$ for small values of $R$, indicating a preference for newly-mobile sites to arise close to previously-mobile sites.  This preference is further quantified by the mobility transfer function,
\begin{equation}
F(t) = \sum\limits_{R'=0}^{R_\mathrm{cut}} P_\mathrm{F}(R',t) \ \ \Big{/} \ \ \sum\limits_{R'=0}^{R_\mathrm{cut}} P^\mathrm{u}_\mathrm{F}(R',t) .
\end{equation}
We use $R_\mathrm{cut} =  2$ lattice sites, but we obtain similar results for $0 < R_\mathrm{cut} < 4$.  The inset of Fig.~\ref{fig:fig2}a shows the time evolution of $F(t)$ as a function of temperature.  We find that, as for atomistic and molecular glass formers, $F(t)$ exhibits a peak value, $F^{\mathrm{max}}(T)$, that shifts to later times and grows with decreasing temperature.  Fig.~\ref{fig:fig2}a shows that $F^{\mathrm{max}}(T)$ is roughly proportional to the mean distance between excitations $\ell$ for the east model.  {This indicates that the probability of encountering nearby mobile sites at random rather than through facilitation decreases proportionally with $\ell$, or equivalently, $c^{-1/\mathrm{d}}$.}

We find that this relationship breaks down when softness is added to the system.  In particular, Fig.~\ref{fig:fig2}b shows that softened systems exhibit a crossover, where $F^{\mathrm{max}}(T)$ first increases and then decreases as a function of temperature.  This crossover in $F^{\mathrm{max}}(T)$ corresponds to a crossover in the relaxation mechanism, where soft delocalized dynamics -- given an Arrhenius temperature dependence for our models -- becomes more probable than DF, which has a super-Arrhenius temperature dependence.  {This crossover behavior is generic for any scenario wherein soft relaxation exhibits a weaker temperature dependence than facilitated dynamics.}  Thus, the temperature variation of $F^{\mathrm{max}}(T)$ gives a qualitative measure of softness, provided that soft delocalized relaxation becomes dominant at lower $T$.  The fact that atomistic and molecular simulated glass formers do not exhibit such a crossover~\cite{vogel2004spatially, bergroth2005examination} indicates that soft delocalized relaxation does not dominate the dynamics, at least over the range of temperatures studied.  We find that finite size effects, which appear when $\ell$ approaches the order of the system size, give rise to a qualitatively similar crossover behavior, and therefore care should be taken when interpreting simulation or experimental results in future studies.

\subsection{Avalanches}

\begin{figure}
\centerline{\includegraphics[width=0.9\columnwidth]{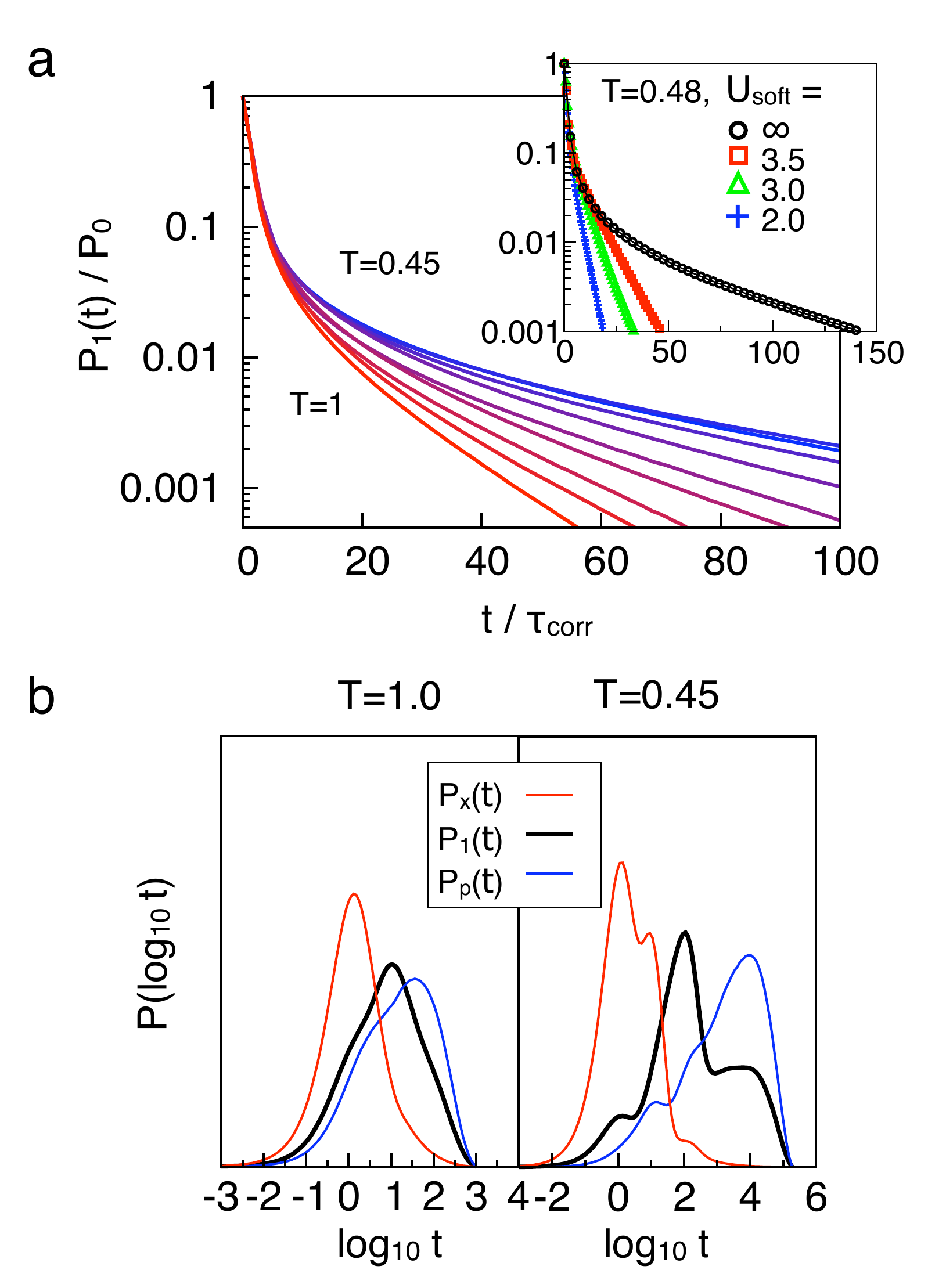}}
\caption{\label{fig:fig3} { Facilitation waiting time distributions (a) Probability distribution of lag times between adjacent kinks, $P_1(t)$, as a function of temperature for the east model.  The inset shows the effect of including soft, non-localized relaxation for one temperature.  The parameter $P_o$ corresponds to the value of the first bin of the $P_1(t)$ histogram.  (b)  Exchange and persistence time distributions $P_\mathrm{x}(t)$ and $P_\mathrm{p}(t)$ for the east model at two different temperatures compared to $P_1(t)$.}}
\end{figure}

Candelier, Dauchot, Biroli and co-workers~\cite{candelier2009building, candelier2010spatiotemporal, candelier2010dynamical} have extensively characterized the dynamics of so-called ``cage-escapes,'' defined by particles that obtain a new center of vibrational motion.  Cage escapes coalesce into clusters in space and time that resemble the excitation dynamics described in Ref.~\cite{keys2011excitations}.  The tendency for clusters of cage escapes to facilitate one another is characterized by a distribution of waiting times between adjacent clusters, $\tau_1$.  {For KCMs, $\tau_1$ is a waiting time between a kink at given lattice site and the next kink at any neighboring site.  This is expressed mathematically for a given lattice site $i$ according to,
\begin{equation}
\tau_1 =  \underset{ j \in \mathrm{\{nbrs}(i)\} }{\operatorname{\min}} \left\{ \underset{t \in [\delta t, \infty) }{\operatorname{arg\min}} \left\{-(1/t)(1-\delta [ \kappa_i(0) \kappa_{j} (t) ] ) \right \} \right\}.
\label{eq:tau1}
\end{equation}
Here, $\{ \text{nbrs}(i) \}$ is the set of all sites $j$ satisfying $D(i,j) = 1$.}  In simulated supercooled liquids and granular materials in experiment, the probability distribution of $\tau_1$, $P_1(t)$, resembles the superposition of  two exponential distributions~\cite{candelier2009building}.  The time constant $\tau_\mathrm{corr}$ characterizes short-time exponential behavior.  It is speculated that this duality of $\tau_1$ distributions implies two distinct physical mechanisms, with short lag times arising from facilitated dynamics and long lag times arising from soft delocalized dynamics~\cite{candelier2010spatiotemporal, candelier2010dynamical}.  This argument implies that the east model, or any other purely-facilitated KCM, should exhibit single-exponential behavior in $P_1(t)$, since DF is the only relaxation mechanism in these systems.  However, as illustrated in Fig.~\ref{fig:fig3}a, the distribution of $\tau_1$ for the east model resembles the double-exponential behavior observed for model liquids and granular materials.  Moreover, the inset of Fig.~\ref{fig:fig3}a shows that introducing soft, delocalized dynamics tends to diminish the two-exponential effect, rather than amplify it.  

Our findings for the east model highlight the important distinction between excitations and their dynamics.  While excitations necessarily connect throughout space and time for facilitated models, their dynamics becomes intermittent at lower $T$, giving rise to a wide range of lag times between events.  This can be rationalized in terms of persistence and exchange times, quantities that arise in the context of transport decoupling in supercooled liquids~\cite{jung2005dynamical, hedges2007decoupling}.  In KCMs, persistence times, $\tau_\mathrm{p}$, are the time scales over which randomly-chosen lattice sites exhibit their first kink.  Exchange times, $\tau_\mathrm{x}$ are the lag times between subsequent kinks at the same lattice site~\cite{jung2005dynamical}.  Like the exchange and persistence time distributions, $P_\mathrm{x}(t)$ and $P_\mathrm{p}(t)$, the distribution of facilitation lag-times $P_1(t)$ exhibits rich behavior, and is described only approximately by the sum of two exponentials, at least for KCMs.  This is demonstrated by plotting $P_1(t)$ on a log-linear scale, as depicted in Fig.~\ref{fig:fig3}b.  The figure shows that $P_1(t)$ involves a combination of persistence and exchange-like time scales.  As $P_\mathrm{x}(t)$ and $P_\mathrm{p}(t)$ decouple at low temperatures~\cite{jung2005dynamical, hedges2007decoupling}, $P_1(t)$ becomes more separated between long and short lag times.  This may explain the tendency for clusters of cage escapes to form large avalanches at higher temperatures, but become more sporadic at low temperatures, when clusters are grouped into avalanches based on the short-time exponential time scale of $P_1(t)$, as described in Ref.~\cite{candelier2010dynamical}.

\begin{figure*}[ht]
\centerline{\includegraphics[width=1.05\textwidth]{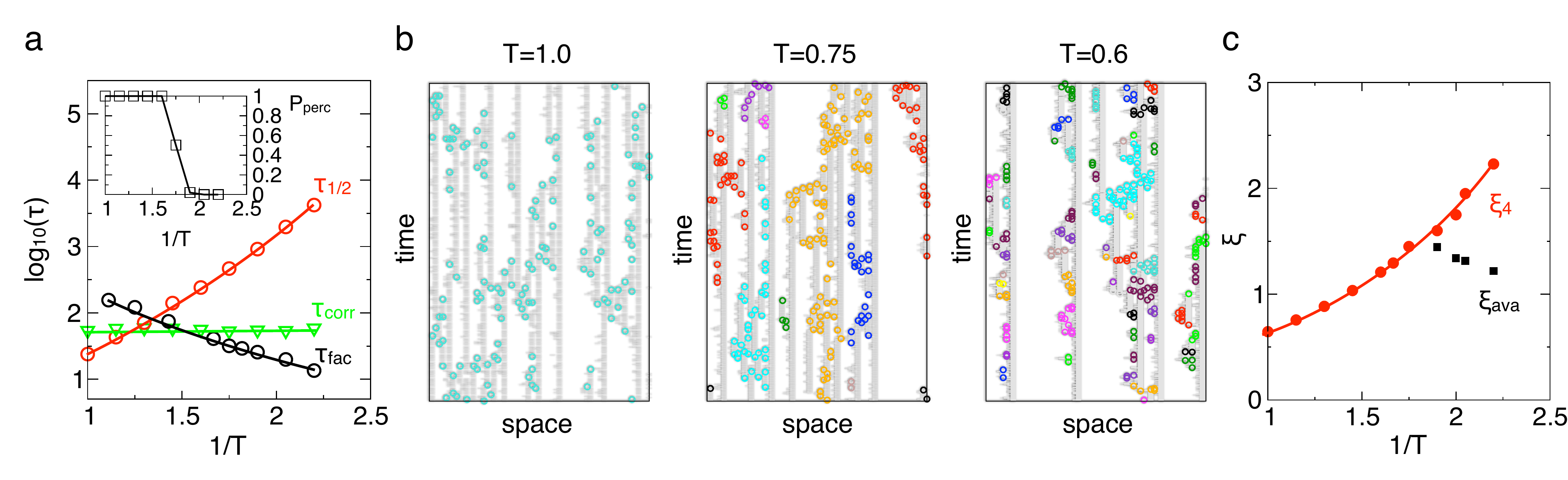}}
\caption{\label{fig:fig4} { Avalanches in KCMs (a) Time scales $\tau_\mathrm{corr}$, $\tau_\mathrm{fac}$ and $\tau_{1/2}$ as a function of temperature for enduring kinks in the east model (see text for definitions).  The inset shows the probability of observing a spanning avalanche of time duration $\tau_{1/2}$ or greater as a function of temperature.  (b)  Avalanches in the east model at three different temperatures.  Enduring kinks are depicted by circles and colored according to the avalanche to which they belong.  Trajectories span approximately $3 \tau_{1/2}$. (c) The four point correlation length $\xi_4$ and the avalanche spatial length scale $\xi_\mathrm{ava}$ as a function of temperature.}}
\end{figure*}

We explore this possibility further by explicitly defining avalanches for the east model.  In the analysis described above, we considered waiting times for kinks $\kappa_i(t)$ to place emphasis on the fact that the observed double-exponential behavior in $P_1(t)$ is robust, even for unprocessed dynamics.  In the analysis that follows, we consider what we believe to be a more realistic mapping of KCMs to clusters of cage escapes by replacing kinks $\kappa_i(t)$ with enduring kinks $h_i(t)$ in Eq.~\eqref{eq:tau1}.  This does not change the qualitative behavior of $P_1(t)$, but rather truncates the distribution for very short lags, less than the sojourn time, $\tau_1 < t_\mathrm{s}$.  Avalanches are defined according to Refs~\cite{candelier2009building, candelier2010spatiotemporal, candelier2010dynamical} by grouping enduring kinks that are adjacent in both space and time.  Space-time is discretized into points $(i, t)$ with temporal lattice spacing $\delta t$.  Points $(i, t)$ and $(j, t')$ belong to the same avalanche $a$ if they contain enduring kinks separated by $\left| t-t' \right| < \tau_\mathrm{corr}$ and $D(i, j)  < r_\mathrm{corr}$.  Each point belongs to exactly one avalanche by construction.  
The cutoff $r_\mathrm{corr}$ is taken to be two lattice sites and $\tau_\mathrm{corr}$ is set by the short-time exponential cutoff of $P_1(t)$ for enduring kinks.  Because $P_1(t)$ is only approximately exponential, as described above, $\tau_\mathrm{corr}$ depends somewhat on the histogram bin size, $\Delta \tau_1$.  For $\Delta \tau_1$ sufficiently small ($ \Delta \tau_1 \approx 10$) and chosen consistently for all data sets, the qualitative behavior of avalanches is not affected.  

Fig.~\ref{fig:fig4}a compares $\tau_\mathrm{corr}$ to the average time duration of an avalanche, $\tau_\mathrm{fac}$, defined by,
 {
\begin{equation}
\tau_\mathrm{fac} = \left<  \underset{ (i, t) \in a \ , \ (j, t') \in a  }{\operatorname{\max}}\left\{ t - t' \right\} \right>_A .
\end{equation}
Here, $\left< \cdots \right>_A$ denotes an ensemble average over avalanches, $a$ at a given state point.}  For comparison, Fig.~\ref{fig:fig4}a also shows $\tau_{1/2}$, the average minimum time required for $1/2$ the lattice sites to exhibit a kink,
{
\begin{equation}
\tau_{1/2} = \left< \underset{t'}{\operatorname{arg\min}} \left\{ \mathcal{P}(t; t')  \leq 0.5 \right\} \right> .
\end{equation}
Here, $\mathcal{P}(t; t')$ is the persistence function~\cite{jung2005dynamical} for a trajectory spanning times $t$ through $t+t'$,
\begin{equation}
\mathcal{P}(t; t')  = \frac{1}{N} \sum_{i=0}^N \delta \left[ \sum_{t''=t}^{t + t'} \kappa_i^2(t'') \right].
\end{equation}}
The time scale $\tau_{1/2}$ is similar to $\tau_\alpha$ for particulate systems~\cite{candelier2010dynamical}.  While $\tau_\mathrm{corr}$ does not vary much with temperature, $\tau_\mathrm{fac}$ decreases as $T$ is lowered, in agreement with the results obtained for the air-driven granular system studied in Ref.~\cite{candelier2010dynamical}. 
 
The sequence of trajectories in Fig.~\ref{fig:fig4}b illustrates the qualitative evolution of avalanches with changing temperature.  At high $T$, dynamics occurs within a single avalanche.  As temperature is lowered, avalanches become increasingly intermittent and spatially separated.  This is quantified by measuring the probability of observing a spanning avalanche with $\tau_\mathrm{fac}  > \tau_{1/2}$ as a function of $T$.  As shown in the inset of Fig.~\ref{fig:fig4}a, the spanning probability crosses over sharply at around $T=0.6$.  This crossover behavior is robust, although we find that the position of the crossover depends on the parameters chosen to define the avalanches.  This indicates that a decreasing spatiotemporal extent of avalanches with supercooling does not preclude DF but rather seems to represent an intrinsic behavior of purely-facilitated systems.  

While the temperature variation of avalanche time scales closely mirrors that of molecular simulations and granular materials, the variation of spatial length scales is more difficult to compare.  Close inspection of the lowest temperature state point depicted in Fig~\ref{fig:fig4}b reveals that the average spatial extent of the avalanches becomes smaller than the typical dynamical length scale $\ell$ (the distance between excitations, shown in grey).  This is quantified by computing the mean end-to-end distance of an avalanche~\cite{candelier2010dynamical},
 {
\begin{equation}
\xi_\mathrm{ava} = \left<  \underset{ (i, t) \in a \ , \ (j, t') \in a }{\operatorname{\max}}\left\{ D(i, j) \right\} \right>_A .
\end{equation}
}
The avalanche length scale is compared to the dynamical correlation length $\xi_4$\cite{chi4, berthier2005numerical}, defined in terms of,
\begin{equation}
\chi_4(t) = N \left[  \left< q(t)  \right>^2 - \left< q(t)^2 \right> \right].
\end{equation}
The indicator function $q_i(t) = \delta \left[ \sum_{t'=0}^t  \kappa^2_i(t') \right]$ is zero if at least one kink has occurred at lattice site $i$ over a time window $t$ and one otherwise.  We obtain $\xi_4$ from the peak value of $\chi_4(t)$, which depends on temperature, $\xi_4 = \left[ \chi_4^\mathrm{max}(T) \right]^{1/d} $.   We opt to define the four-point length scale dynamically for closer correspondence with particulate systems; however, $\ell$ is directly proportional to $\xi_4$ and thus the quantities are interchangeable.

At high temperatures, for which avalanches span the system, we find that $\xi_\mathrm{ava}$ exceeds $\xi_4$.  This behavior is also observed for the granular system studied in Ref.~\cite{candelier2010dynamical}.  There, it is argued that large avalanches on length scales exceeding $\xi_4$ result from a collection of dynamically independent events occurring on a scale $\xi_\mathrm{ava} \approx \xi_4$.  Following this argument, we estimate $\xi_\mathrm{ava} \approx \xi_4$ in this range.  Below a crossover temperature, $\xi_\mathrm{ava}$ becomes smaller than $\xi_4$ and decreases with supercooling.  The low dimensionality of the model studied here may accentuate this trend, but this qualitative behavior should hold in general.  This stands in contrast with the interpretations of Ref~\cite{candelier2010dynamical}, where it is speculated that avalanches span a typical dynamical correlation length at all state points, and thus represent dynamically-independent events.  The available data for particulate systems seems to neither prove nor disprove this hypothesis.  Ref~\cite{candelier2010dynamical} finds that $\xi_\mathrm{ava} \approx \xi_4$ near the crossover, but this is true by construction, given that a crossover exists and the temperature variation of $\xi_\mathrm{ava}$ is not strong, as is the the case for the east model.  Evidence of dynamically-independent avalanches would involve $\xi_\mathrm{ava}$ that grows in proportion to $\xi_4$ over a range of supercooled state points, but, for the narrow range of state points available for particulate systems, $\xi_\mathrm{ava}$ neither definitely grows nor definitively shrinks.  We thus leave our findings as predictions to be investigated through future study.

By studying avalanches in KCMs, we arrive at a somewhat different interpretation regarding the specific relationship between avalanches and DF than that previously postulated.  In particular, we find that the observed phenomenology previously thought to be inconsistent with DF -- long waiting times between avalanches and the decreasing spatiotemporal size of avalanches with supercooling -- follows from facilitated models and does not imply either long-ranged correlations between excitations~\cite{candelier2010spatiotemporal} or a breakdown of DF at low temperatures or high packing fractions~\cite{candelier2010dynamical}.  Despite these differences, our interpretations are in agreement with the overarching spirit of Refs.~\cite{candelier2009building, candelier2010spatiotemporal, candelier2010dynamical} -- that is, we find that avalanches do represent a type of facilitated dynamics and that localized dynamics in particulate systems do not map to \emph{directly} to excitations in KCMs.  Our findings reinforce the idea that avalanches are a robust phenomenon warranting future study, particularly with regard to their dependence on temperature and dimensionality, as well as their dynamic independence and connection to excitation dynamics.

\subsection{Facilitation Volume}

The mobility transfer function and avalanches described in the previous sections involve the practical difficulty of defining cutoff values that, if chosen incorrectly, can affect the qualitative outcome of the measurements.  Ref. ~\cite{keys2011excitations} introduces an alternate, parameter-free measure of DF known as the facilitation volume, which quantifies the overall impact of an initial rearrangement on the subsequent mobility field~\cite{keys2011excitations}.  The time-dependent displacement field conditioned on an enduring kink at a tagged site $1$ is given by,
\begin{equation}
\tilde{\mu}(r, t, t') =  \frac{\left< h_1(0) \sum\limits_{j=1}^N \mu_j(t, t') \delta \left[ r - \left| D(1, j) \right| \right]  \right> }{ \left<  h_1(0) \right> n(r) }.
\end{equation}
Here $r$ is an integer in the range $[0, N/2]$ and $n(r)$ is the probability density of observing a distance $r$ between lattice sites, $n(r) = \left[1 + \Theta(r) \right]$.  We consider the behavior of $\tilde{\mu}(r, t, t')$  over a time window spanning from $\Delta t/2$ to $t$, where $\Delta t/2$ demarcates the completion of the enduring kink at the origin.  Thus, the initial dynamics does not factor into the value of the displacement field.  The quantity $\tilde{\mu}(r, t, t')$ is closely related to the distinct contribution to the four-point susceptibilities $\chi_4$ and $\chi_U$, often used to characterize DH~\cite{chi4, donati1999spatial}; however, $\tilde{\mu}(r, t, t')$ focuses specifically on dynamical correlations with initial dynamics, whereas four-point functions are sensitive to large-scale dynamical correlations that build up from the initial dynamics over time.  That is, $\tilde{\mu}(r, t, t')$ is a three-point correlation function~\cite{dalle2007spatial}, rather than a four-point correlation function.

\begin{figure}[h]
\centerline{\includegraphics[width=0.85\columnwidth]{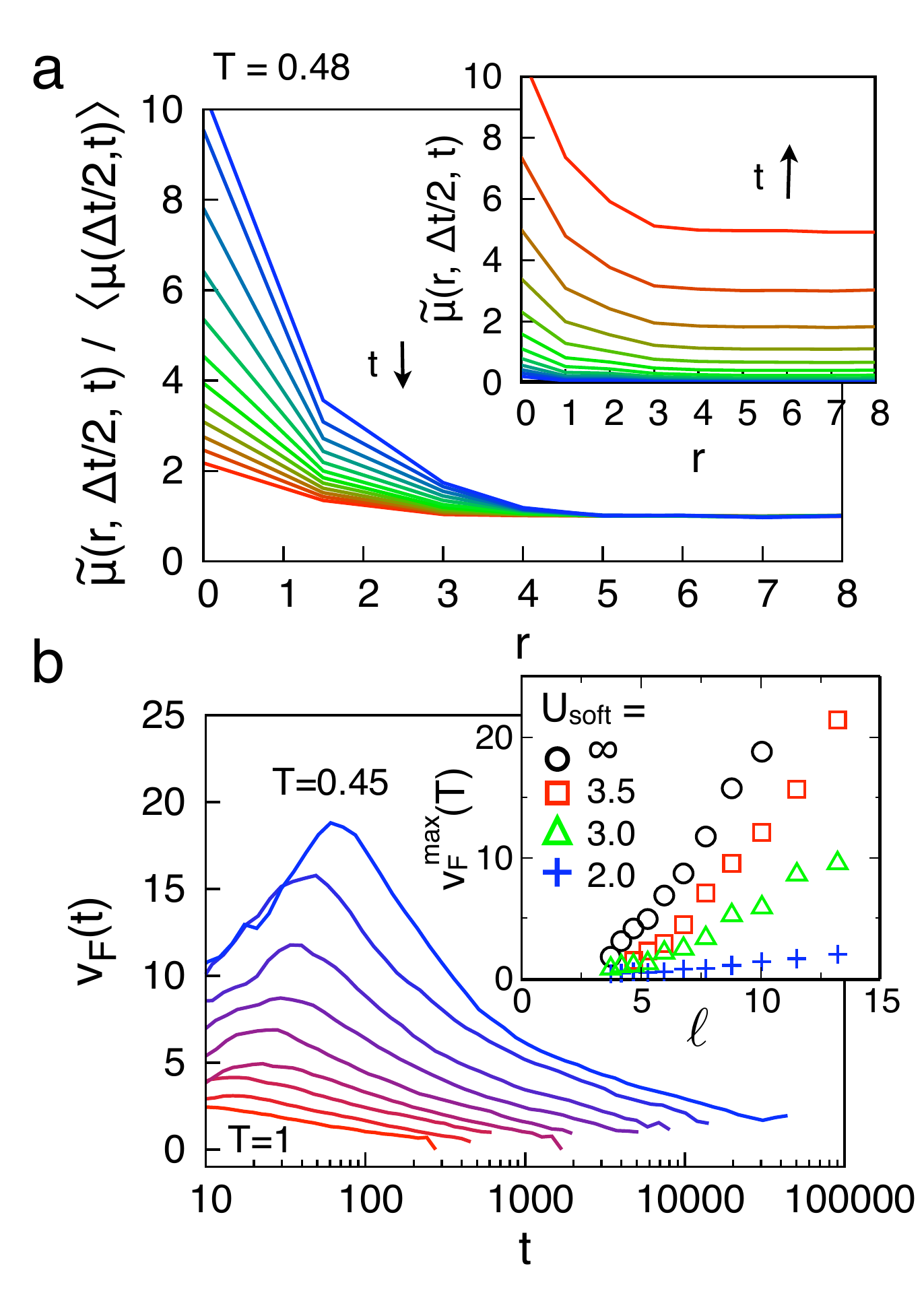}}
\caption{\label{fig:fig5} { Mobility field and facilitation volume.  (a) Time evolution of the mobility field $\tilde{\mu}(r, \Delta t/2, t)$ for the east model for a single temperature, $T=0.48$.  For both the main panel and inset, the time window $t$ spans from $t \approx t_s$ (blue) to $t \approx 5 \tau$ (red).  (b) Facilitation volume $v_\mathrm{F}(t)$ as a function of temperature.  Temperatures span the same range as in Fig.~\ref{fig:fig2}a.  The inset shows the peak value of $v_\mathrm{F}(t)$ at each temperature $v_\mathrm{F}^{\mathrm{max}}(T)$ for different values of softness, controlled by $U_\mathrm{soft}$.   }}
\end{figure}

The behavior of $\tilde\mu(r, \Delta t /2, t)$ is plotted in Fig.~\ref{fig:fig5}a as a function of the time window $t$ for the east model.  The inset shows that, for $t_s \leq t \leq 5\tau$, $\tilde\mu(r, \Delta t /2, t)$ peaks near position of the previously-enduring kink at the origin, and decays to $\left< \mu(0, t) \right>$ as $r$ becomes large.  The main panel shows the same curves normalized by $\left< \mu(0, t) \right>$, highlighting the excess displacement relative to the uncorrelated result obtained in absence of the initial dynamics.  The curves plateau at $r \approx \ell/2$, indicating that mobility correlations span a maximum range $\ell \approx 1/c$.  The fact that the length scale is relatively constant for a large range of time scales reflects the non-zero probability of observing a chain of excitations that spans $\ell$, even on very short time scales.

The facilitation volume is defined as the sum of the excess mobility over all $r$,
\begin{equation}
v_\mathrm{F}(t) = \sum_{r=0}^{N/2} \left[ \frac{ \tilde{\mu}(r, \Delta t / 2, t) } { \left < \mu(\Delta t/2,t) \right> } - 1 \right].
\end{equation}
The behavior of $v_\mathrm{F}(t)$ is plotted in Fig.~\ref{fig:fig4}b as a function of temperature.  For each temperature, the function peaks near the sojourn time $t_s$ and decays toward zero as $t$ becomes much greater than $\tau$.  This is similar to the behavior observed for particulate models, exception that, for those systems, $v_\mathrm{F}(t)$ peaks near the $\alpha$ relaxation time~\cite{keys2011excitations}.  This discrepancy may arise due to the non-discrete nature of particulate systems, where vibrational motions dominate the displacement field at very short times.  Regardless of the characteristic time chosen in the range $t \leq \tau$, $v_\mathrm{F}(t)$ grows with decreasing temperature, and, in particular, scales as $\ell$.  This is illustrated by the inset of Fig.~\ref{fig:fig5}b, which shows the peak of $v_\mathrm{F}(t)$ as a function of temperature, $v_\mathrm{F}^{\mathrm{max}}(T)$.  Due to statistical uncertainty, it is unclear whether the facilitation volumes reported for atomistic systems in Ref.~\cite{keys2011excitations} scale approximately with $\ell^{d}$, but this relationship should be explored in more detail in the future.

Fig.~\ref{fig:fig5}b also shows the variation of $v_\mathrm{F}^{\mathrm{max}}(T)$ with the characteristic energy for soft delocalized relaxation, $U_\mathrm{soft}$.  Allowing for non-facilitated dynamics moves adds background noise to the displacement field, which diminishes the excess mobility and reduces $v_\mathrm{F}^{\mathrm{max}}(T)$ accordingly.  For large amounts of softness, relaxation occurs mostly in the absence of facilitation, and $v_\mathrm{F}^{\mathrm{max}}(T)$ no longer exhibits significant growth with decreasing $T$.  In contrast to $F^\mathrm{max}(T)$, the value of $v_\mathrm{F}^{\mathrm{max}}(T)$ does not exhibit a temperature crossover for any value of $U_\mathrm{soft}$.  This is related to the fact that $v_\mathrm{F}^{\mathrm{max}}(T)$ is normalized by $\left< \mu(\Delta t /2, t) \right>$, which decreases as a function of $T$, whereas $F^{\mathrm{max}}(T)$ is normalized by the same function for all $T$.  Thus, our findings imply that large values of $v_\mathrm{F}^{\mathrm{max}}(T)$ that grow with supercooling, such as those observed in Ref~\cite{keys2011excitations}, indicate the presence of significant DF, but do not rule out the possibility of some delocalized relaxation as well.

\section{Discussion}

Our results indicate that the current body of literature regarding DF implies that DF is present at all state points investigated, and becomes increasingly apparent with increased supercooling.  It remains possible that DF could be superseded by another mechanism at currently-inaccessible conditions, but evidence for such a mechanism has not yet been reported.  {The fact that structural relaxation data for a wide range of experimental conditions collapses~\cite{elmatad2009corresponding} to a universal functional form~\cite{elmatad2009corresponding, garrahan2003coarse, heuer2008exploring, heuer2008properties} seems to indicate that the relaxation mechanism does not change.  Still, interpretations involving an explicit crossover within the supercooled regime are possible~\cite{xia2001microscopic}.}

Our results regarding avalanches imply that clusters of cage escapes are closely related to excitation dynamics and might be applied to study DF in the future, particularly for experimental systems, or other systems where inherent states or time-coarse graining on a very fine scale is not possible.  In Ref.~\cite{candelier2010spatiotemporal} soft modes~\cite{widmer2008irreversible} were invoked to explain the sudden triggering of avalanches at points in space with little prior dynamical activity over long periods of time.  Although our results imply that such long waiting times are a natural consequence of facilitated dynamics, the observed connection with soft modes implies that these quantities might provide a generic method for detecting excitations in absence of their dynamics~\cite{ashton2009relationship}.  

In addition to the quantities involving DF explored here, the method that we derive for a displacement field for KCMs might allow for additional detailed mapping between the dynamics of KCMs and molecular systems.  The quantities reported in Ref~\cite{keys2011excitations} seem like a particularly fruitful avenue for future study.

\section{Acknowledgements}
YSE was supported by an NSF GRFP fellowship during the initiation of the project and by New York University's Faculty Fellow program during the later stages.  ASK was supported by Department of Energy Contract No. DE-AC02Ñ05CH11231. We thank D. Chandler for his guidance and insight.  We thank T. Speck, D.T. Limmer, G. D\"uring, and E. Lerner for helpful comments regarding the manuscript.  Without implying either agreement or disagreement with our interpretations, we thank A. Widmer-Cooper, D.R. Reichman, G. Biroli, O. Dauchot, and R. Candelier for constructive correspondences regarding this work.


\end{document}